\documentclass[conference]{IEEEtran}
\IEEEoverridecommandlockouts
\usepackage{cite}
\usepackage[nolist,nohyperlinks]{acronym}
\usepackage{amsmath,amssymb,amsfonts}
\usepackage{algorithmic}
\usepackage{graphicx}
\usepackage{url}
\usepackage{textcomp}
\usepackage{tikz}
\usepackage{adjustbox}
\usepackage{pgfplots}
\usepgfplotslibrary{statistics} 
\usepackage{xcolor}
\def\BibTeX{{\rm B\kern-.05em{\sc i\kern-.025em b}\kern-.08em
    T\kern-.1667em\lower.7ex\hbox{E}\kern-.125emX}}
\pgfplotsset{compat=1.16} 

\newcommand\copyrighttext{%
  \footnotesize © 2021 IEEE.  Personal use of this material is permitted.  Permission from IEEE must be obtained for all other uses, in any current or future media, including reprinting/republishing this material for advertising or promotional purposes, creating new collective works, for resale or redistribution to servers or lists, or reuse of any copyrighted component of this work in other works.}
\newcommand\copyrightnotice{%
\begin{tikzpicture}[remember picture,overlay]
\node[anchor=south,yshift=10pt] at (current page.south) {\fbox{\parbox{\dimexpr\textwidth-\fboxsep-\fboxrule\relax}{\copyrighttext}}};
\end{tikzpicture}%
}

\begin{document}

\title{A Voting-Based Blockchain Interoperability Oracle\\
}



\author{
 \IEEEauthorblockN{
    Michael Sober\IEEEauthorrefmark{1}\IEEEauthorrefmark{2},
    Giulia Scaffino\IEEEauthorrefmark{1}\IEEEauthorrefmark{3},
    Christof Spanring\IEEEauthorrefmark{4},
    Stefan Schulte\IEEEauthorrefmark{1}\IEEEauthorrefmark{2}}
 \IEEEauthorblockA{\IEEEauthorrefmark{1}\textit{Christian Doppler Laboratory for Blockchain Technologies for the Internet of Things} \\
 \IEEEauthorrefmark{2}\textit{Institute of Data Engineering, TU Hamburg, Hamburg, Germany} \\
    \{michael.sober, stefan.schulte\}@tuhh.de}
 \IEEEauthorblockA{\IEEEauthorrefmark{3}\textit{Institute of Logic and Computation, TU Wien, Vienna, Austria} \\
    giulia.scaffino@tuwien.ac.at}
 \IEEEauthorblockA{\IEEEauthorrefmark{4}\textit{Pantos GmbH, Vienna, Austria} \\
    christof.spanring@bitpanda.com}
}

\newpage

\maketitle

\begin{abstract}

Today's blockchain landscape is severely fragmented as more and more heterogeneous blockchain platforms have been developed in recent years. These blockchain platforms are not able to interact with each other or with the outside world since only little emphasis is placed on the interoperability between them. Already proposed solutions for blockchain interoperability such as naive relay or oracle solutions are usually not broadly applicable since they are either too expensive to operate or very resource-intensive.

For that reason, we propose a blockchain interoperability oracle that follows a voting-based approach based on threshold signatures. The oracle nodes generate a distributed private key to execute an off-chain aggregation mechanism to collectively respond to requests. Compared to state-of-the-art relay schemes, our approach does not incur any ongoing costs and since the on-chain component only needs to verify a single signature, we can achieve remarkable cost savings compared to conventional oracle solutions.

\end{abstract}


\copyrightnotice

\begin{acronym}
	\acro{IoT}{Internet of Things}
	\acro{SPV}{Simplified Payment Verification}
	\acro{zkSNARK}{Zero-Knowledge Succinct Non-Interactive Argument of Knowledge}
	\acro{TLS}{Transport Layer Security}
	\acro{HTTPS}{Hypertext Transfer Protocol Secure}
	\acro{SGX}{Software Guard Extensions}
	\acro{HTLC}{Hashed Timelock Contract}
	\acro{BLS}{Boneh-Lynn-Shacham}
	\acro{DKG}{Distributed Key Generation}
	\acro{VSS}{Verifiable Secret Sharing}
	\acro{SHA}{Secure Hash Algorithm}
	\acro{EVM}{Ethereum Virtual Machine}
	\acro{RPC}{Remote Procedure Call}
	\acro{BAR}{Byzantine-Altruistic-Rational}
	\acro{ECDSA}{Elliptic Curve Digital Signature Algorithm}
	\acro{gRPC}{gRPC Remote Procedure Calls}
	\acro{HTLC}{Hashed Timelock Contract}
\end{acronym}
\section{Introduction}
\label{sec:introduction}

In recent years, blockchain technology has become increasingly important not only as the underlying technology for cryptocurrencies~\cite{nakamoto2008bitcoin} but also in many other application areas including supply chain management~\cite{saberi2019blockchain}, healthcare~\cite{aguiar2020health}, and others. This has led to the development of more and more heterogeneous blockchain platforms~\cite{schulte2019towards, belchior2020survey}. These are often tailored to specific requirements, as there is not one blockchain that is capable of fulfilling the (often) disjunctive needs of different application areas.

Research and industry tend to not consider the interoperability between different blockchain platforms, turning these platforms into self-contained systems~\cite{schulte2019towards}. As a result, these platforms are not able to collaborate to benefit from novel features and properties of other or newly developed blockchain platforms. The problem of weak interoperability does not only exist within the world of blockchains but generally with systems that are outside the blockchain platform's boundaries. As an example, we can use the smart contract and decentralized application platform Ethereum~\cite{wood2014ethereum}. Smart contracts running in the \ac{EVM} cannot interact with the outside world, which means that a smart contract is only able to read and modify the state of the hosting blockchain. For many applications, however, it is important to be able to obtain data from the outside world. One can imagine, for example, a decentralized application that needs flight data from an airline to determine if a user has the right to compensation in case a flight is delayed or a cross-chain token transfer that requires the knowledge of whether the tokens were burned on the original blockchain. This problem is also known as the blockchain oracle problem~\cite{al2020trustworthy}.

Many attempts have been made to achieve interoperability between different blockchain platforms~\cite{belchior2020survey} and to solve the oracle problem~\cite{heiss2019oracles, al2020trustworthy}. One possible approach to achieve interoperability between two blockchain platforms is the use of blockchain relays~\cite{buterin2016chain}. Such relays are usually provided as smart contracts running on a ``target blockchain'', i.e., the blockchain which needs data from a ``source blockchain''. With blockchain relays, the state of one blockchain is replicated within another blockchain, which enables callers of the relay contract to verify the existence of a transaction on the source blockchain.

Bitcoin and Ethereum use Merkle trees to store transactions in a block, whereby the root of the Merkle tree is stored in the block header. Therefore, a relay contract can use \ac{SPV} which utilizes Merkle proofs to check whether a transaction is included in a block of a source blockchain~\cite{buterin2016chain}. Since block header validation is done by a smart contract on-chain, no trusted intermediary is needed to relay the block headers. However, this has the disadvantage that such relays also have considerable costs, since on-chain verification is very expensive and blocks have to be relayed continuously, even if they are perhaps not needed at all.

Another approach is to make use of blockchain oracles to get information from the outside world. Blockchain oracles are bridges between blockchain platforms and external data sources. The task of a blockchain oracle is to query data from external data sources and then to pass the data items to a smart contract. The problem here is to ensure the authenticity and integrity of the data because we have to trust the oracle that it behaves honestly~\cite{heiss2019oracles}.

One can differentiate between oracles that are based on a centralized or a decentralized approach~\cite{al2020trustworthy}. Centralized oracles represent a single point of failure since trust in a single oracle node is required. Following the decentralized model, there is no single point of failure and the trust assumption moves from one oracle node to multiple oracle nodes. Many decentralized oracles follow a voting-based approach to provide data, i.e., the votes are aggregated to determine the overall result. Unfortunately, the aggregation mechanism can become very expensive, if carried out on-chain.

To make a step towards blockchain interoperability and overcome the issues of current relay schemes and oracle solutions, we investigate the application of \ac{BLS} threshold signatures to create an off-chain aggregation mechanism to reduce the operating costs of the oracle. Further, we examine the use of \ac{DKG} protocols to generate distributed private keys which are necessary for the creation of the threshold signatures while also preserving the decentralized nature of the blockchain. We provide the system design of a voting-based blockchain interoperability oracle that makes use of the aforementioned concepts and enables clients to verify that a transaction is included in another blockchain. Further, we deliver a prototypical implementation and show the applicability of our proposed solution by conducting a security and cost analysis.

The remainder of this paper is organized as follows: In Section~\ref{sec:background}, we introduce some basic concepts needed in our approach. Subsequently, we present the design of the oracle in Section~\ref{sec:system_design}. We follow up with implementation details of the prototype in Section~\ref{sec:implementation} and evaluate our solution in Section~\ref{sec:evaluation}. Afterward, we discuss related work in Section~\ref{sec:related_work}. Finally, Section~\ref{sec:conclusion} concludes the paper.
\section{Background}
\label{sec:background}
In the course of this section, we explain the basics of \ac{BLS} signatures, followed by a short description of \ac{VSS} which leads us to a discussion of \ac{DKG} protocols.

\subsection{\ac{BLS} Signatures}
\label{sec:bls_signatures}

The \ac{BLS} signature scheme~\cite{boneh2004short} is based on elliptic curve pairings ($e: G_1 \times G_2 \rightarrow G_T $) respectively bilinear maps. Using BLS, a public key $PK$ is generated by multiplying the selected private key $SK$ with the generator $G$ of a cyclic group. To create a signature $\sigma$ one has to hash the message $m$ on the curve $H(m)$. This can be accomplished by using a hashing algorithm like the \ac{SHA}-256, whereby the resulting hash is used as the x-coordinate of a point on the curve. 

If it is not possible to find such a point using this x-coordinate, it can simply be incremented until a valid point is found. This is an essential difference from other signature schemes in which the hash can be used directly. After that, the signature can be calculated by multiplying the point with the private key. To verify the signature, it comes down to checking the bilinear pairings as can be seen in~Eq.~\ref{eq:pairing_check}.

\begin{equation}
\label{eq:pairing_check}
    e(\sigma, G) = e(H(m), PK)
\end{equation}

This signature scheme has the advantage that it allows to generate particularly short signatures. Another very important aspect especially for this work is that it is possible to create threshold signatures~\cite{boldyreva2002threshold} using a secret sharing scheme~(see~Section~\ref{sec:verifiable_secret_sharing}), whereby multiple participants share a distributed private key and $t$ out of $n$ signature shares are required to create a valid signature. 

Also, it enables the aggregation of multiple signatures~\cite{boneh2003aggregate}, whereby we only need to verify two elliptic curve pairings to verify the aggregate signature. This is particularly interesting in the area of blockchain technology. For example, one could instead of verifying every single signature of the transactions in a block, only verify a single aggregate signature.


\subsection{Verifiable Secret Sharing}
\label{sec:verifiable_secret_sharing}

With the help of a secret sharing scheme, it is possible to divide a secret $S$ into $n$ shares whereby each party gets a different share of $S$, but at least $t$ shares need to be known to recover $S$. Otherwise, it is not possible to get any knowledge about $S$. These schemes are also known as $(t,n)$ threshold schemes. 

One of the first secret sharing schemes has been proposed by Shamir~\cite{shamir1979share} in 1979. Using Shamir's scheme, a dealer picks a polynomial $f(x)$~(see~Eq.~\ref{eq:random_polynomial}) of degree $t-1$ whereby the constant term $a_0$ of the polynomial is equal to $S$ and the coefficients are chosen randomly.

\begin{equation}
\label{eq:random_polynomial}
    f(x) = a_0 + a_1x + \dots + a_{t-1}x^{t-1}
\end{equation}

The secret can be shared with $n$ different parties by evaluating the polynomial at the positions $x_n={1..n}$ and distributing the values for $f(x_n)$ along with $x_n$ to the respective parties. Since every polynomial of degree $t-1$ is defined by exactly $t$ points, we can make use of polynomial interpolation (e.g., using the Lagrange interpolation formula) to recover the original polynomial and to evaluate it at position $0$ to get the secret. 

One problem with this, however, is that Shamir's scheme does not take into account that a dealer could distribute incorrect or inconsistent shares. Furthermore, the shareholders could also return incorrect shares. We cannot assume that the dealer and the shareholders are trustworthy, which is why we need secret sharing schemes that also take this kind of behavior into consideration, as is the case with \ac{VSS} schemes. By using such schemes, each party can verify that it has received the correct information from the dealer and that shareholders submitted the correct shares. Hereby, we limit ourselves mainly to non-interactive VSS schemes, in which only the dealer sends messages and no communication between the other participants is necessary. This reduces the communication overhead.

One commonly used example of such a scheme is Feldman's \ac{VSS}~\cite{feldman1987practical}. Feldman's \ac{VSS} is based on Shamir's secret sharing scheme but additionally makes use of homomorphic encryption to commit to the secret and coefficients of the random polynomial. The commitments are broadcast while the shares are distributed through private channels. Each party can use the commitments to verify the validity of its share, due to the homomorphic property of the used encryption scheme.

A problem with this approach is that the commitment to the secret also leaks information about the secret. This is where Pedersen's \ac{VSS}~\cite{pedersen1991non}, another very well-known scheme, constitutes a more secure solution. Like Feldman's \ac{VSS}, it is also based on Shamir's secret sharing scheme, but it makes use of a different commitment scheme that ensures that no information about the secret is leaked unless one can find a solution to the discrete logarithm problem.

\subsection{Distributed Key Generation}
\label{sec:distributed_key_generation_bg}

In the previous section, we discussed \ac{VSS}, whereby such schemes ensure that a dealer distributes the secret correctly and the shareholders cannot provide incorrect shares of the secret. However, the dealer yet has to be trusted as it still knows the secret. This is, among other things, a big concern in the area of threshold cryptography, where we want that only certain subsets of $t \leq n$ participants can jointly encrypt data or create a signature. Since the dealer knows about the distributed private key, it has the opportunity to encrypt data or to create signatures without the consent of at least $t$ out of $n$ participants.

To avoid this, a \ac{DKG} protocol that allows the generation of distributed private keys without the dependency on any trusted third party can be used. In such protocols, no party owns the private key and the private key is never reconstructed. Many such protocols have been proposed over time~\cite{pedersen1991threshold, gennaro1999secure, kate2009distributed, kokoris2020asynchronous}, whereby we limit ourselves to the first proposed \ac{DKG} protocol by Pedersen~\cite{pedersen1991threshold} to describe the basic concept. 

In Pedersen's \ac{DKG} protocol, every participant acts as a dealer during one of the $n$ parallel executions of Feldman's \ac{VSS} scheme to share a randomly picked secret. Each participant publishes its commitments using a broadcast channel, such as a blockchain. After that, every participant sends the signed private shares to the other participants through private channels. When receiving a share, each participant verifies the share by using the previously published commitments. If it is an invalid share, a complaint including the share and the signature is broadcast. Each participant computes its private share by summing up all shares received from the other parties which shared their secret correctly and computes the public key through the published commitments.
\section{System Design}
\label{sec:system_design}

In this section, we propose the design of a voting-based blockchain interoperability oracle. Initially, we give a brief overview of the basic concept. We follow up with a description of the architecture and finally provide a comprehensive definition of the functionality.

\subsection{Overview}
\label{sec:overview}

\begin{figure}[t]
  \centering
  \begin{adjustbox}{width=0.73\linewidth}
  \includegraphics[width=\linewidth]{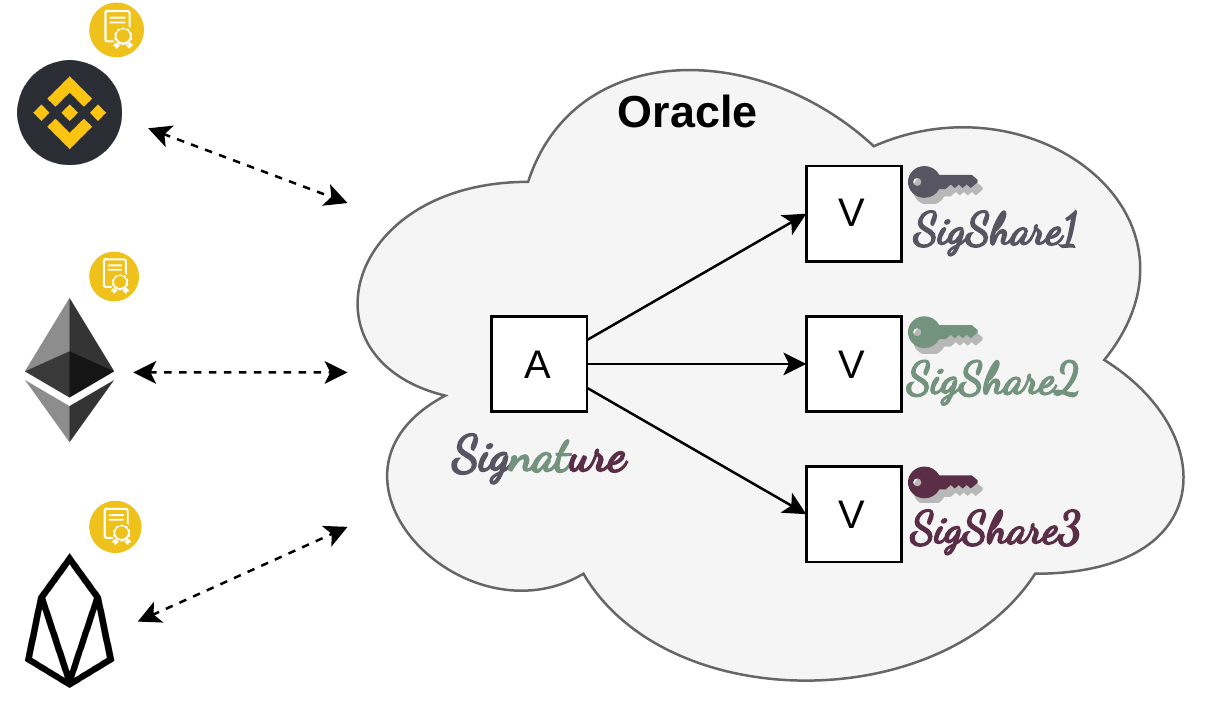}
  \end{adjustbox}
  \caption{Overview of the System}
  \label{fig:system_overview}
\end{figure}

The proposed design for a voting-based blockchain interoperability oracle~(see~Figure~\ref{fig:system_overview}) allows clients to verify that a transaction is included in another blockchain. It uses an off-chain aggregation mechanism~(see~Section~\ref{sec:offchain_aggregation}) based on \ac{BLS} threshold signatures. The oracle nodes are divided into one aggregator and multiple validators which can collectively respond to requests issued by clients (i.e., parties which are interested in the verification of a transaction from another blockchain). For this, the oracle nodes use a \ac{DKG} protocol to generate a distributed private key~(see~Section~\ref{sec:distributed_key_generation}), whereby each node only knows its private key share and the shared public key. Validators obtain the data from the other blockchain and sign it with their private key share, while the aggregator collects the data and the signature shares from the validators to recover the final signature to submit it along with the data to the smart contract. If the aggregator is not able to collect at least the threshold of signature shares with the same result, it cannot generate a valid signature. To ensure reliability, the aggregator changes over time so that each oracle node takes over the task of the aggregator at some point.

As an integral part of the system, we also apply an incentive mechanism~(see~Section~\ref{sec:incentive_mechanism}) to encourage oracle nodes to engage as part of the decentralized oracle and to behave honestly. For this, the client must provide compensation for the transaction fees that arise from submitting the result using the oracle contract and also offer two additional rewards. These additional rewards consist of the aggregation reward and the validation reward. Without these rewards, the aggregator, as well as the validators, would have no interest in participating as they would only be providing their resources without getting anything back.

\begin{figure*}[t]
  \centering
  \begin{adjustbox}{width=0.71\linewidth}
  \includegraphics[width=\linewidth]{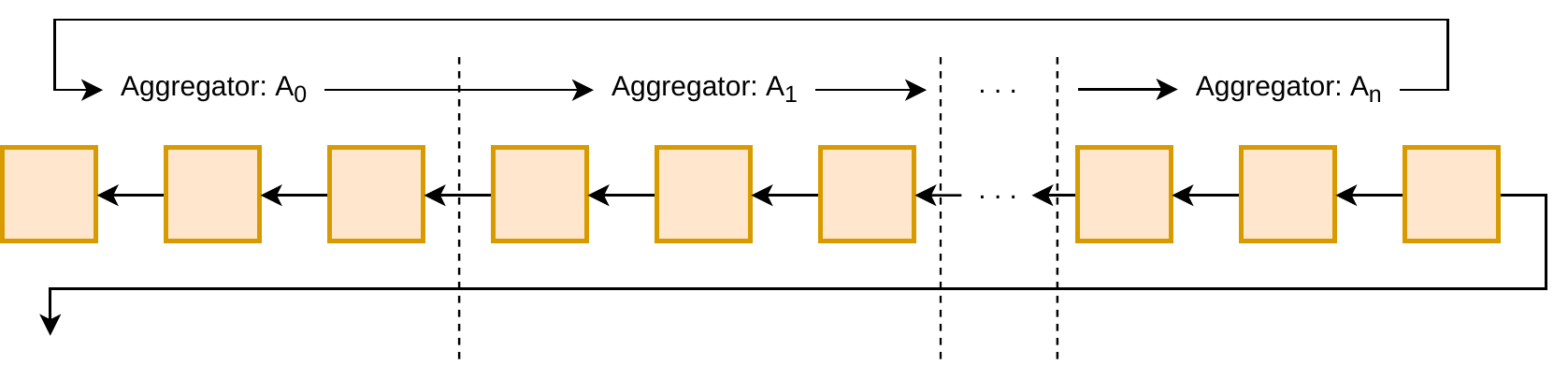}
  \end{adjustbox}
  \caption{Aggregator selection mechanism}
  \label{fig:aggregator_selection}
\end{figure*}

\subsection{Architecture}
\label{sec:architecture}

In our architecture, we can differentiate between the components that are on the blockchain, i.e., the smart contracts, and the components that are off-chain, i.e., the oracle nodes. 

For the on-chain components, we make use of three different smart contracts. The first of these is a registry contract, which is responsible for managing all oracle nodes~(see~Section~\ref{sec:oracle_registration}) and selecting the current aggregator~(see~Section~\ref{sec:offchain_aggregation}). Via this smart contract, an oracle node discovers the other oracle nodes and checks if they got selected as the current aggregator. 

The next component is the oracle contract, which receives requests from clients and sends them to all oracle nodes. Furthermore, the oracle contract is also responsible for receiving and verifying the responses from the aggregator~(see~Section~\ref{sec:offchain_aggregation}) and transferring the rewards. Further, it also stores all responses and makes them available to the clients. 

The last smart contract is the key contract, which is responsible for managing the public key and initiating the generation of new keys~(see~Section~\ref{sec:distributed_key_generation}). The off-chain component consists of oracle nodes that can take on the tasks of the aggregator as well as the validator. 

\subsection{Oracle Registration}
\label{sec:oracle_registration}

The first step of an oracle node to join the decentralized oracle is to register using the registry contract. During registration, the oracle nodes must provide their host address and \ac{BLS} public key. Furthermore, a stake must be deposited which is used as a measure against Sybil attacks~(see~Section~\ref{sec:sybil_attacks}). After an oracle node has registered, it is eligible to take part in the run of the \ac{DKG} protocol.

Another important point is that oracle nodes must also be able to deregister. Furthermore, oracle nodes can also be kicked out through a majority vote if they behave incorrectly and lose their stake.

Particular care is taken to ensure that the system remains fully operational. A distinction is made between the signature threshold and the threshold of qualified validators. The validator threshold is required to ensure that the system continues to work even if several validators fail or are misbehaving. Should the number of validators fall below the validator threshold, a new key generation process is triggered.

\subsection{Distributed Key Generation}
\label{sec:distributed_key_generation}

Each time a new oracle node registers, the registry contract checks whether a new run of the \ac{DKG}~protocol should be initiated. For this, we set a certain number, which indicates how many new registrations are necessary. Should the number be reached, the registry contract calls the key contract to trigger the start of the \ac{DKG} protocol. The key contract then broadcasts a generation event containing the threshold which should be used.

On receiving the generation event, oracle nodes need to wait a certain amount of time (e.g., measured in blocks) to ensure that every oracle node had a chance to receive the event since it can take some time for a block to get propagated through the network. Then, the oracle nodes execute the \ac{DKG} protocol with all currently registered oracle nodes. In our approach, we make use of Pedersen's \ac{DKG} protocol (see Section~\ref{sec:distributed_key_generation_bg}). However, other \ac{DKG} protocols could also be used, as long as the run of the protocol is made transparent through the usage of blockchain technology.

The entire run of the \ac{DKG} protocol takes place off-chain, which means that we need to get the generated public key as well as the number of qualified validators into the blockchain, i.e., we already have the oracle problem in our proposed solution. One possibility of getting the public key into the blockchain is to use an on-chain aggregation mechanism. However, this approach has the disadvantage that it gets more expensive the higher the number of oracle nodes becomes. These costs could be neglected if the frequency of generating new keys is very low. The same applies to the possibility of using a dispute mechanism, which is used by our prototype whereby only one oracle node submits the shared public key and the other oracle nodes can dispute the key. Finally, one could also assume that several other oracles already exist that can be used for this task. These could be oracles that may be pursuing a completely different approach, or it is already another instance of the proposed oracle solution. With the latter, however, we have a bootstrapping problem where the first instance cannot call an existing oracle. Therefore, one would have to use one of the first two approaches or a centralized solution and change it later on.

\subsection{Off-chain Aggregation}
\label{sec:offchain_aggregation}

\begin{figure*}[t]
  \centering
  \begin{adjustbox}{width=0.73\linewidth}
  \includegraphics[width=\linewidth]{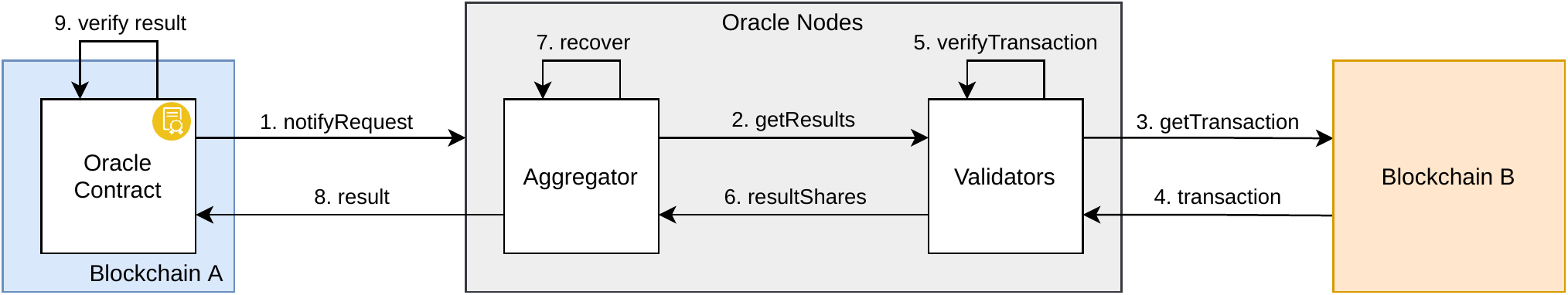}
  \end{adjustbox}
  \caption{Aggregation mechanism}
  \label{fig:aggregation_mechanism}
\end{figure*}

The task of aggregating the results and the signature shares is taken over by an aggregator, which is one of the oracle nodes. The aggregator changes in a cycle of \textit{n} blocks based on a round-robin mechanism~(see~Figure~\ref{fig:aggregator_selection}). This has several advantages over an approach in which anyone can aggregate and submit the voting results: It prevents multiple simultaneous submissions of which only the first one would be successful, while all the other oracle nodes which also try to submit an aggregation result have to pay the costs for the failed transaction without getting compensated. This would lead to the problem that oracle nodes are not incentivized to submit a result for which it is not certain whether the submission would be successful or not. Since this is relatively difficult to determine, nodes could become reluctant to act as an aggregator.

Another benefit is that this approach also consumes less bandwidth since not all oracle nodes try to aggregate a result, which reduces the number of exchanged messages. The fact that the aggregator changes over time, still ensures that if an aggregator should fail, the next aggregator will take its place and normal operation can be continued.

The aggregation mechanism~(see~Figure~\ref{fig:aggregation_mechanism}) starts when a client sends a request to the oracle contract. After that, all oracle nodes are notified about the request~(Step~1). Subsequently, the aggregator begins to collect results from the validators~(Step~2), whereby the request only needs to contain the request number so that the validators know which request should be fulfilled. On receiving a request from the aggregator, a validator retrieves the transaction from the target blockchain~(Step~3~and~4), verifies the transaction~(Step~5), and returns the response consisting of the response to the request and the signature share~(Step~6).

The aggregator collects results until it has at least threshold \textit{t} identical results with valid signature shares. If the aggregator does not receive at least $t$ results with valid signature shares, it will not be able to recover the signature. This can happen, as there can be temporary inconsistencies in blockchain systems, validators fail or behave incorrectly. 

Regarding the first problem, however, one can assume that consistency will be achieved at some point (depending on the source blockchain) and that the threshold of the validators will agree. This is one of the advantages that is given by the fact that this oracle is limited to providing data from other blockchains. In this case, the aggregator can try again after a while, or if the aggregator changes in the meantime, the new aggregator will take over the request and repeat the voting.

If the aggregator has received enough shares, it recovers the full \ac{BLS} signature using Lagrange interpolation~(Step~7). After recovering the full \ac{BLS} signature, the aggregator can submit the result to the oracle contract by providing the result and the signature~(Step~8). The oracle contract then hashes the result to a point on the curve and checks the elliptic curve pairing~(Step~9). If the elliptic curve pairing is successful, the smart contract calculates the reward~(see~Section~\ref{sec:incentive_mechanism}) and transfers it to the aggregator. Finally, the oracle contract notifies the client that the result is available. Here, however, one could also pursue a different approach in which the client also defines a callback function which is to be called when the result gets submitted. In this case, other problems would have to be considered, such as the costs caused by the callback function. Furthermore, the result is then not easy to obtain for other clients.

\subsection{Incentive Mechanism}
\label{sec:incentive_mechanism}

Since the aggregator only submits a response including the data and the threshold signature, the oracle contract does not have any information about the validators who were contacted, which makes it difficult to reward validators directly. We cannot rely on the aggregator to provide us with this information either, as it cannot be verified whether it is correct, as the entire \ac{DKG} protocol is carried out off-chain. Furthermore, it is not particularly practical to reward each validator individually, as this would result in very high fees for the client, not only because the client has to pay each validator individually but also because each transfer causes additional costs during the execution of the smart contract.

Therefore, we propose to only reward the aggregator for submitting the result. The aggregator gets compensated for the transaction costs, receives an aggregation reward, and additionally has the chance to win the validation reward with a certain probability. The probability is scaled super-linearly based on the deposited stake to encourage the creation of only one identity~\cite{cai2020oracle}. If the aggregator is not lucky and does not win the validation reward, the reward is maintained by the oracle contract until an aggregator gets lucky enough to win the accumulated validation rewards. 

It is particularly important to ensure that the validators are not able to predict whether the aggregator will win the validation reward. Otherwise, validators would benefit from not providing the aggregator with an answer to be able to increase their own chances of winning when they become aggregators themselves (see Section~\ref{sec:offchain_aggregation}). Therefore, we leverage the unpredictable randomness provided by the signature which is recovered by the aggregator to decide whether the aggregator receives the validation reward. In the case of the aggregator, it does not matter whether it knows if it will receive the validation reward before submitting, since the aggregator is still encouraged to submit the result, as to at least receive the aggregation reward. As for the validators, they are still incentivized to answer the current aggregator with the assumption that it will not receive the reward and thus increase their winnable reward when it is their turn to be an aggregator.

Since a round-robin mechanism is used for aggregator selection, care should be taken that the chance of winning the validation reward is not too high. Validators who were recently selected as the aggregator have a reason to assume that another aggregator will win the validation reward during the time they will have to wait to be selected again. They may be tempted to stop validating until they believe otherwise. Therefore, the chance should be small enough such that the validators can still assume that no one will win the reward in the current aggregation round.
\section{Implementation}
\label{sec:implementation}

After defining the system design of our solution, we created a prototypical implementation. Our prototype enables Ethereum-based blockchains to exchange data with each other. However, the solution can be implemented to work with other blockchains. For this, the target blockchain needs to have smart contract capabilities and enable elliptic curve pairing checks. In this section, we discuss the used technologies as well as the implementation of the smart contracts and the oracle node which is available as open-source software on GitHub\footnote{\url{https://github.com/pantos-io/ioporacle}}.

\subsection{Smart Contracts}

For the implementation of the prototype, we decided in favor of Ethereum because it is one of the most popular second-generation blockchains and accordingly a wide range of tools is available. Another important factor was that Ethereum already provides a precompiled contract that enables elliptic curve pairing checks on the alt\_bn128 curve.

The registry contract stores all oracle nodes in an iterable mapping and provides clients with the necessary functions to retrieve them. For the aggregator selection mechanism, we defined a threshold of six blocks after which the aggregator will be switched. Further, after every third registration, the registry contract calls the key contract to trigger a new execution of the \ac{DKG} protocol.

On receiving the call from the registry contract, the key contract calculates the threshold based on the number of currently registered oracle nodes and emits the key generation event. For the threshold, we defined that the majority of all registered nodes is necessary to produce a valid signature. For submitting the key, we provide a function in the prototype via which the public key can be set by one of the oracle nodes. The public key can be disputed if the majority of oracle nodes agree.

When implementing the oracle contract, we especially paid attention to implementing the contract in a gas-efficient way. Since storage operations are among the most expensive, we have tried to keep the number of these low. If a client sends a request to the oracle contract, the request is only emitted as an event and not saved in storage. This is possible because only oracle nodes need to be able to read the requests. However, this approach involves more work for the oracle nodes since they have to filter the blockchain for past events in case they missed some of them.

As has already been mentioned before, we use a precompiled contract that allows pairing checks for the alt\_bn128 curve to verify the \ac{BLS} signatures submitted by the aggregator. These precompiled contracts are already existing contracts that run outside the \ac{EVM} and perform more complex tasks. 
One of the advantages of these contracts is that they are usually cheaper. Further, we make use of the try and increment approach~(see~Section~\ref{sec:background}) to hash the response on the curve.

\subsection{Oracle Node}

The oracle node is implemented in the Go programming language. For the implementation, we used the advanced crypto library Kyber\footnote{\url{https://github.com/dedis/kyber}}. We adapted the library to work with the alt\_bn128 curve used by Ethereum. The library provides the necessary packages for threshold \ac{BLS} signatures and the implementation of a \ac{DKG} protocol which is based on the protocol proposed by Pedersen~(see~Section~\ref{sec:distributed_key_generation_bg}). For the \ac{DKG} protocol, we needed a broadcast channel, for which we opted for the IOTA tangle~\cite{popov2018tangle}, as it enables fee-less and publicly verifiable data exchange. Other broadcast channels (e.g., Ethereum) can also be used, but they may incur additional costs. The oracle nodes use a Go client to create transactions and send them to an IOTA node. These are zero-value transactions that are only used to exchange messages which are necessary to execute the \ac{DKG} protocol.

Furthermore, the oracle nodes must be able to communicate directly with one another. The aggregator must be able to collect the results from the validators and all oracle nodes need a private channel to each other to distribute the private shares during the execution of the \ac{DKG} protocol. Therefore, we make use of \ac{gRPC} to connect the oracle nodes. 

To interact with the smart contracts and retrieve data from an Ethereum blockchain, the oracle nodes need access to an Ethereum node whereby we make use of the Go Ethereum client to connect to the \ac{RPC} server. Further, we use a Go binding generator to create the counterparts of the smart contracts in Go to produce as little boilerplate code as possible.
\section{Evaluation}
\label{sec:evaluation}

In this section, we analyze the security and the costs of the proposed solution. This provides insight into if the solution is applicable, what problems can arise and what needs to be considered.

\subsection{Security Analysis}
\label{sec:security}

To analyze the security of the oracle, we look at various attack scenarios and the consequences that can result from them. In particular, we are looking at lazy voting, free loading, Sybil attacks, and the key submission.

We can categorize the oracle nodes based on the \ac{BAR} model proposed in~\cite{aiyer2005bar}, which has already found application in other works for security analysis in the area of blockchain technology, e.g.,~\cite{frauenthaler2020eth}: \emph{Rational} oracle nodes deviate from the protocol as long as they will be able to increase their benefits by doing so. \emph{Byzantine} oracle nodes, however, can unexpectedly deviate from the protocol for unknown reasons, whereby it does not matter if it is intentional or unintentional misbehavior and whether they gain a benefit from it. \emph{Altruistic} oracle nodes will always adhere to the protocol no matter if the rational choice would provide them with additional benefits.

The proposed oracle solution applies an incentive mechanism~(see~Section~\ref{sec:incentive_mechanism}) to encourage all oracle nodes to follow the protocol. However, it should be mentioned that even if all incentives are aligned properly, rational oracle nodes may deviate from the protocol. The reason for this is that also actors outside the oracle have to be taken into account. Hence, it may seem rational for oracle nodes to follow the protocol solely based on the applied incentive mechanism, but external factors can influence their decision by providing better benefits for arbitrary reasons.

\subsubsection{Lazy Voting}
 
In the lazy voting problem, rational oracle nodes do not deliver the correct result but always provide the same response to maximize their benefits. If e.g., it is the case that a certain result occurs particularly often, it can be more beneficial to always return the same result directly instead of executing the request. This could lead to an incorrect result being successfully submitted. In our proposed solution, this would mean that a lazy validator would always respond to the request ``Is transaction $tx$ included in blockchain $B$ and confirmed by at least $n$ blocks?'' with $true$. Since it is more important for many use cases that a certain transaction is included and confirmed, it can be assumed that requests will be answered more frequently with $true$ than with $false$. However, this problem can be circumvented by modifying the request. We expand the usual request and ask ``In which block on blockchain $B$ is transaction $tx$ included and is it confirmed by at least $n$ blocks?'' instead. This question forces lazy validators to read from blockchain B since otherwise, they cannot know in which block the transaction is contained. The lazy voting problem is also discussed in related work~(see~Section~\ref{sec:oracles}) about decentralized oracles.

\subsubsection{Free Loading}

For the creation of a valid signature, the aggregator only needs to collect as many valid signature shares as the threshold $t$ specifies. This means that in the best case only $t$ validators have to execute the request. However, all other validators who have not contributed to the creation of the signature also have the chance to win the validation reward, when selected as an aggregator.

The problem here is that some validators may never respond for the aforementioned reason. Even though rational oracle nodes may be able to increase their benefits by not responding, they should still be encouraged to respond. The reason for this is that they minimize the risk of the aggregator not being able to get enough responses, which leads to a lower possible validation reward. Even if more than $t$ validators behave altruistically, it may be the case that no signature can be created due to possible inconsistencies. This indicates that it is more beneficial for the validators to always respond to requests. Above all, validators do not have to perform any particularly resource-intensive tasks, which means that possible resource savings are low. 

\subsubsection{Sybil Attacks}
\label{sec:sybil_attacks}

Another important aspect that must be considered is to what extent Sybil attacks are possible and what the consequences are. A Sybil attack describes the threat that single faulty participants can control multiple identities, which enables them to compromise larger parts of a system. Douceur~\cite{douceur2002sybil} has shown that it is not possible to prevent such attacks without a central authority except for conditions that are not practicable for large-scale distributed systems. Since we are in the area of blockchain technology where we usually do not have a central authority, we have to consider such attacks. Further, in permissionless blockchains, everyone is allowed to join the system, whereby it is an easy task to create new identities by simply generating a new key pair. The same applies to the proposed oracle solution where any number of oracle nodes can register.

We have already mentioned in Section~\ref{sec:system_design} that oracle nodes have to deposit a stake to make Sybil attacks more difficult. The intention is to make the creation of new identities expensive so that it becomes more difficult to control larger parts of the oracle. On the other hand, we also make it more difficult for honest oracle nodes to join the oracle. Therefore, the trade-off for a more Sybil-resistant oracle is less decentralization considering only oracle nodes that can provide the stake can participate. Nevertheless, an attacker is still able to register multiple oracle nodes to be able to create more signature shares, which enables the attacker to gain more voting power.

In the worst case, an attacker could register threshold $t$ or more oracle nodes and decide on the result alone if it can provide the necessary stake. To provide better Sybil resistance, we further encourage the creation of only one identity, by increasing the chance of winning the validation reward super-linearly based on the deposited stake. As a result, it is more beneficial for oracle nodes to create a single identity with a higher stake than to split the stake between multiple identities, as this gives them a greater chance of winning the validation reward. Therefore, if one is only interested in gaining more benefits, this approach resembles the rational choice.

\subsubsection{Key Submission}

The selected key submission mechanism also imposes some security risks. As has already been discussed, the public key can be submitted by using an on-chain aggregation mechanism, dispute mechanism, or another oracle solution for the key submission~(see~Section~\ref{sec:distributed_key_generation}). While the use of an oracle solution appears to be more cost-effective, we must be aware of the security risks that arise from this approach.

A central oracle solution represents a single point of failure, which can submit arbitrary public keys to then be able to create valid signatures. As a result, a single participant could gain full control over the oracle. However, if one uses a decentralized approach, the risk is lower.

It must be ensured that both oracles, i.e., the oracle used for the key submission and the interoperability oracle, are as independent of each other as possible. Otherwise, oracle nodes that participate in both oracles would be able to change the result in their favor. In the worst case, this would mean that a certain subset of oracle nodes could have complete control over the oracle used for key submission and can thus select the public key. This means that even if the attacker is unable to create more than threshold $t$ oracle nodes in the interoperability oracle, it can get control over the interoperability oracle if it can control the oracle for key submission.

\subsection{Cost Analysis}
\label{sec:costs}

We analyze the costs of the proposed solution by comparing the implemented prototype with two alternative approaches which do not make use of \ac{BLS} threshold signatures. Accordingly, we implement two additional oracle contracts (\textit{On-chain Oracle}, \textit{ECDSA Oracle}) which follow different aggregation mechanisms. Furthermore, we compare the costs of our approach with the costs incurred by ETH Relay~\cite{frauenthaler2020eth}, a novel relay scheme~(see~Section~\ref{sec:oracles}), to examine how well our approach performs compared to state-of-the-art schemes. To ensure repeatability, the implementations, as well as the evaluation scripts, are also included in the open-source project on GitHub (see Section~\ref{sec:implementation}).

The \textit{On-chain Oracle} implements an on-chain aggregation mechanism whereas each oracle node calls the oracle contract to submit a result. The \textit{ECDSA Oracle} makes use of \ac{ECDSA} signatures to verify the result. In contrast to our proposed scheme, an aggregator does not submit a single \ac{BLS} signature but rather submits several \ac{ECDSA} signatures that are verified by the oracle contract. Since there is no reasonable source of randomness for either of these variants, the reward is paid out to each oracle node that is part of the majority.

For the experiment, we use a private Ethereum blockchain based on the Muir Glacier hard fork, on which we deploy all smart contracts. In this experiment, we request the verification of a transaction with every type of the aforementioned aggregation mechanisms. Besides, we are changing the number of participating oracle nodes to be able to determine how the costs develop with an increasing amount of oracle nodes.

By comparing the different mechanisms~(see~Figure~\ref{fig:cost_comparison}), it can be seen that the costs for the on-chain aggregation mechanism are considerably higher than those of the other two. This is due to the reason that for the on-chain mechanism more storage space is needed and each participant has to create a transaction to submit its vote. In comparison, the other mechanism in which an aggregator only submits several \ac{ECDSA} signatures, is more cost-efficient. The problem here, however, is that the costs continue to rise with the number of oracle nodes, even though the verification of \ac{ECDSA} signatures is a relatively cheap operation on the Ethereum platform.

\begin{figure}[t]
\centering
\begin{adjustbox}{width=0.58\linewidth}
\begin{tikzpicture}
\begin{axis}[
    boxplot/draw direction = y,
    xtick={1},
    xticklabels={BLS Result Submissions},
    ylabel=Gas Consumption
]
\addplot+[
boxplot prepared={
median=255779,
upper quartile=270834.5,
lower quartile=245712.5,
upper whisker=357977,
lower whisker=230979,
}, fill, fill opacity=0.5, draw=black] coordinates {};
\end{axis}
\end{tikzpicture}
\end{adjustbox}
\caption{Gas consumption of the result submission with BLS signatures}
\label{fig:cost_submission}
\end{figure}
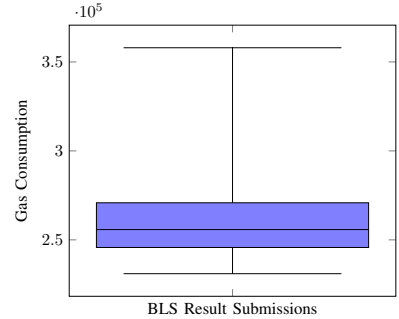

Our proposed solution based on \ac{BLS} threshold signatures, on the other hand, causes almost constant costs that are independent of the number of oracle nodes, since only one signature has to be verified. The costs are only almost constant as the try and increment approach is used for hashing. As can be seen in Figure \ref{fig:cost_submission}, submitting the result consumes on average 257,607 gas with a standard deviation of 21,671 gas. Verifying a \ac{BLS} signature is an expensive operation, but it is considerably more cost-efficient as the number of oracle nodes increases. With more than three oracle nodes, it is already cheaper than the on-chain mechanism and with more than 15 nodes, it is also cheaper than the \ac{ECDSA} mechanism. Therefore, by using this approach we can achieve a higher degree of decentralization without increasing the costs.

\begin{figure}[t]
\centering
\begin{adjustbox}{width=0.58\linewidth}
\begin{tikzpicture}
\begin{axis}[
    no markers,
    ymax=500000,
    xmax=33,
    ylabel=Gas Consumption,
    xlabel=\# of Oracle Nodes,
    legend style={at={(1.0, 1.15)}},
    legend columns=-1
]
\addplot table [x=no_nodes, y=gas, col sep=comma, mark=none] {./data/on-chain-cost.csv};
\addlegendentry{On-chain}
\addplot table [x=no_nodes, y=gas, col sep=comma, mark=none] {./data/ecdsa-cost.csv};
\addlegendentry{ECDSA}
\addplot table [x=no_nodes, y=gas, col sep=comma, mark=none] {./data/bls-cost-median.csv};
\addlegendentry{BLS}
\end{axis}
\end{tikzpicture}
\end{adjustbox}
\caption{Gas consumption of the different aggregation mechanisms}
\label{fig:cost_comparison}
\end{figure}
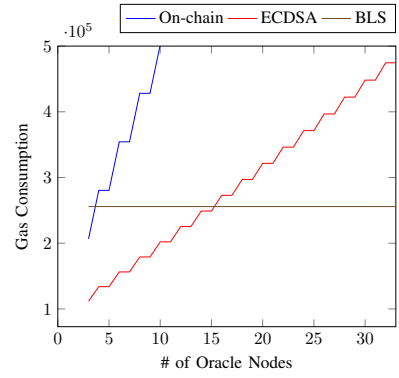

Now that we have compared our approach with two different oracle solutions, we examine how the costs differ compared to a relay solution. To conduct this analysis, we choose ETH Relay because it is an advanced relay solution that is specifically designed to be more cost-effective. For the comparison, we assume a period of 100 blocks. Every block header submission in ETH Relay consumes 284,041 gas with a standard deviation of 3,679 gas. As a result, submitting 100 block headers consumes around 28,404,100 gas. Using our presented approach, we can submit 110 results that incur roughly the same gas costs as the 100 block header submissions of ETH Relay. The costs for the actual request must also be taken into account. In the case of ETH Relay, the relay contract needs to carry out a \ac{SPV} and check the membership of the block in the longest chain, which means that the costs increase as the search depth increases. In the case of our oracle-based relay, however, a request only ever consists of an event being emitted, whereby the result can simply be accessed directly, which results in constant costs.

At the moment, the application of interoperability solutions is a rather infrequent occurrence, which presumably suggests that not many requests are made. In the worst case, this could mean that not a single transaction has to be verified for 100 blocks. With ETH Relay, the blocks still have to be submitted and thus the costs must be sustained. Hence, keeping the relay alive is a huge burden since these submitted block headers do not yield any profit for the submitter. In contrast, our presented oracle-based solution does not incur any costs in this scenario, as every request is fulfilled on demand. However, if the number of requests increases drastically such that it is more than 110, ETH Relay would be the more cost-efficient solution. One must also note that in this case further adjustments can be made to the oracle such that every request enables the verification of all the transactions within a block, rather than a single one, by providing the Merkle root of such a block.
\section{Related Work}
\label{sec:related_work}

So far, different blockchain interoperability solutions have been proposed. These include hash-locks, relay solutions, and oracles. In the course of this section, we examine solutions that are related to our work.

\subsection{Hash-Locks}
\label{sec:hash_locks}

Hash-locks are a well-known technique to enable a basic form of blockchain interoperability without oracles and relays, e.g., to realize atomic cross-chain swaps~\cite{herlihy2018atomic}. Atomic swaps allow multiple parties to exchange their assets across multiple blockchains. The involved parties make use of \acp{HTLC} to escrow their assets with a hashlock~\textit{h} and timelock~\textit{t}. Ownership of the asset is only transferred if the receiver can provide the secret~\textit{s} such that $h(s) = h$ before \textit{t} expires. However, attention must be paid to specify the right timelock values and use the correct deployment order of the contracts.

\subsection{Relays}
\label{sec:relays}

Frauenthaler et al.~\cite{frauenthaler2020eth} propose ETH Relay, a novel relay scheme for Ethereum-based blockchains. A validation-on-demand pattern is used to keep the operating costs low by reducing the number of expensive full block header validations. Instead of validating every block header when it is submitted, off-chain clients have a certain time frame in which they can dispute submitted block headers. 
To make \acp{SPV} more efficient, the authors also optimize the traversal of the blockchain, by jumping from branching point to branching point instead of iterating over the whole data structure. While the authors achieve a remarkable cost reduction over traditional relay solutions, the costs remain quite high.

In~\cite{westerkamp2020zk}, the authors present zkRelay, which is a relay solution that utilizes off-chain computations to validate batches of block headers through the usage of \ac{zkSNARK} proofs. Since these proofs are generated off-chain, the smart contract only needs to be able to verify the proof, removing the necessity of storing and validating every submitted block header, whereby only the last block header of a batch is stored. 
Although the solution offers improvements in the form of scalability and cost optimization, there are tradeoffs in terms of delay and hardware resources. 
Our solution, however, can immediately retrieve data from the source blockchain and has no excessive RAM consumption since no complex proofs are generated.

\subsection{Oracles}
\label{sec:oracles}

Provable~\cite{provable} (formerly known as Oraclize) is a centralized oracle service for various blockchain platforms, e.g., Ethereum and EOS. The Provable blockchain oracle utilizes TLSnotary authenticity proofs~\cite{tlsnotary2014} to attest the authenticity of the data retrieved from the originating source. Within TLSnotary, an auditee can prove to an auditor the authenticity of information retrieved from a Web server that is using the \ac{HTTPS} protocol, by utilizing the features of the underlying \ac{TLS} protocols 1.0 and 1.1.

With Town Crier, the authors of~\cite{zhang2016towncrier} propose another centralized blockchain oracle, which uses the \ac{HTTPS}/\ac{TLS} protocol and additionally utilizes trusted hardware to ensure the authenticity of the data. The implementation uses Intel's \ac{SGX} which allows the execution of a process in a protected address space which guards the process against malicious software running outside of the enclave but also from various hardware attacks. While both of the aforementioned oracle services offer a solution to the oracle problem, the high level of centralization poses a major problem regarding scalability and single points of failure.

In~\cite{ellis2017chainlink}, the authors present ChainLink, a decentralized oracle network. ChainLink offers a reputation-based voting system whereby users can issue queries to the ChainLink smart contracts. Queries are executed by the selected oracle nodes which retrieve the results from different or overlapping sets of data sources. These results are aggregated by a smart contract which is also responsible for the calculation of the outcome. Breidenbach et al.~\cite{breidenbach2021chainlink} further introduce a new off-chain reporting protocol for ChainLink, which however follows a different approach compared to our solution.

Peterson et al.~\cite{peterson2019augur} propose a decentralized oracle and prediction market platform called Augur. Within Augur, users can create prediction markets to get information that is external to the system. Market participants trade shares of those markets and reporters can vote by staking their REP tokens (Augur's native token) on one possible outcome. The reached consensus of reporters is considered as the outcome. Depending on the result, reporters receive a reporting fee from the markets. Augur's incentive mechanism encourages participants to behave honestly to maximize their profits, while misbehaving participants get penalized.

The authors of~\cite{adler2018astraea} propose ASTRAEA, another decentralized voting-based blockchain oracle. Submitters, voters, and certifiers play a voting game to decide on the truth value of boolean propositions. These propositions are added to the system by submitters, who pay fees to receive an answer to the submitted proposition. Voters play a low-risk/low-reward game by depositing a stake to answer a random proposition. Certifiers on the other hand play a high-risk/high-reward game whereby they can choose a proposition to certify but have to place a high stake. 

Merlini et al.~\cite{merlini2019public} propose an extension to this protocol to solve the lazy equilibrium problem, whereby all voters report the same answer on all propositions. The authors describe a paired-question protocol, in which submitters add queries with two antithetic questions, whereby the oracle additionally needs to check if both answers converge to different outcomes. With this approach, the protocol ensures that honest voters receive higher rewards than lazy voters. 

While the approaches discussed above are interesting solutions to the oracle problem, they are not specifically aiming at providing blockchain interoperability and also implement their mechanisms on-chain, which results in high costs.
\section{Conclusion}
\label{sec:conclusion}

Research on closing the gaps between different blockchains has already led to several concepts and solutions. However, these are usually too expensive or very resource-intensive.

To overcome these issues, we propose a voting-based blockchain interoperability oracle that uses an off-chain aggregation mechanism based on \ac{BLS} threshold signatures. The oracle nodes are divided into one aggregator and multiple validators and generate a distributed private key to collectively decide on the result of a request. Validators read the data from the other blockchain and sign it with their private key share. The selected aggregator collects the results and the signature shares from the validators to create a valid signature which is submitted to and verified by the oracle contract. Our evaluation shows that the proposed solution is more cost-efficient than other oracle solutions and also incurs lower costs than state-of-the-art relay schemes depending on the request rate.

In future work, we will investigate how we can improve Sybil resistance and the submission of the shared public key. We will also examine if there is still potential to further reduce the costs by applying other signature schemes and enabling requests such that multiple transactions can be verified.
\section*{Acknowledgment}

The financial support by the Austrian Federal Ministry for Digital and Economic Affairs, the National Foundation for Research, Technology and Development as well as the Christian Doppler Research Association is gratefully acknowledged.

\bibliographystyle{IEEEtran}
\bibliography{refs.bib}

\end{document}